\documentclass[conference]{IEEEtran}
\usepackage{fullpage}
\usepackage{amsmath,amssymb} \interdisplaylinepenalty=2500
\usepackage{cite}

\newtheorem{theorem}{Theorem}

\newtheorem{corollary}[theorem]{Corollary}
\newtheorem{lemma}[theorem]{Lemma}

\newtheorem{example}{Example}

\DeclareMathOperator{\rank}{\sf rank\hspace{0.1em}}

\newcommand{\Fq}{\mathbb{F}_q}
\newcommand{\Fqm}{\mathbb{F}_{q^m}}
\newcommand{\bigO}{\mathcal{O}}

\newcommand{\mat}[1]{\begin{bmatrix} #1 \end{bmatrix}}

\newcommand{\defn}{\textit}  %% use for terms defined inline
\title{Security for Wiretap Networks via\\ Rank-Metric Codes}

\author{
\IEEEauthorblockN{Danilo Silva and Frank R. Kschischang}
\IEEEauthorblockA{Department of Electrical and Computer Engineering, University of Toronto \\
Toronto, Ontario M5S 3G4, Canada, {\{danilo, frank\}@comm.utoronto.ca}}
}

\begin{document}
\maketitle
\thispagestyle{empty}

\begin{abstract}

The problem of securing a network coding communication system against a wiretapper adversary is considered. The network implements linear network coding to deliver $n$ packets from source to each receiver, and the wiretapper can eavesdrop on $\mu$ arbitrarily chosen links. A coding scheme is proposed that can achieve the maximum possible rate of $k=n-\mu$ packets that are information-theoretically secure from the adversary. A distinctive feature of our scheme is that it is universal: it can be applied on top of any communication network without requiring knowledge of or any modifications on the underlying network code. In fact, even a randomized network code can be used. Our approach is based on Rouayheb-Soljanin's formulation of a wiretap network as a generalization of the Ozarow-Wyner wiretap channel of type II. Essentially, the linear MDS code in Ozarow-Wyner's coset coding scheme is replaced by a maximum-rank-distance code over an extension of the field in which linear network coding operations are performed.

\end{abstract}

\section{Introduction}
\label{sec:introduction}

The paradigm of network coding \cite{Ahlswede++2000,Li++2003,Koetter.Medard2003} has provided a rich source of new problems that generalize traditional problems in communications. One such problem, introduced in \cite{Cai.Yeung2002:Secure} by Cai and Yeung, is that of securing a multicast network against a wiretapper adversary.

Formally, consider a multicast network with unit capacity edges implementing linear network coding over the finite field $\Fq$. Each link in the network is assumed to carry a packet of $m$ symbols in $\Fq$. We assume that the maxflow from source to each receiver is at least $n$ and that the network code is feasible for the multicasting of $n$ packets, that is, each receiver is able to recover the $n$ packets originated at the source. Now, suppose there is a wiretapper that can listen to transmissions on $\mu$ arbitrarily chosen links of the network. The secure network coding problem is to design a network code and an outer encoder at the source such that a message can be transmitted from the source to each receiver without leaking any information to the wiretapper (i.e., security in the information-theoretic sense).

The work of Cai and Yeung \cite{Cai.Yeung2002:Secure} shows that a solution to this problem exists if the message consists of at most $k=n-\mu$ packets and $q$ is sufficiently large. Their solution involves changing the network code such that certain security conditions are met and requires a field of size at least $\ell \choose \mu$, where $\ell$ is the number of links in the network. Feldman et al. \cite{Feldman++2004:CapacitySecure} simplified the conditions in \cite{Cai.Yeung2002:Secure} and showed that it is possible to achieve security by carefully designing the outer code, while leaving the network code unchanged. They also show that, if a linear outer code is used and the network topology is arbitrary, then there are instances of the problem where a very large field size is necessary to achieve capacity.

Recently, Rouayheb and Soljanin \cite{Rouayheb.Soljanin2007} have shown that the problem of secure network coding can be regarded as a network generalization of the Ozarow-Wyner wiretap channel of type II \cite{Ozarow.Wyner1984,Ozarow.Wyner1985}. Their observation provides an important connection with a classical problem in information theory and yields a much more transparent framework for dealing with network coding security. In particular, they show that the same technique used to achieve capacity of the wiretap channel II---a coset coding scheme based on a linear MDS code---can also provide security for a wiretap network. Unfortunately, in their approach, the network code has to be modified to satisfy certain constraints imposed by the outer code.

Note that, in all the previous works, either the network code has to be modified to provide security \cite{Cai.Yeung2002:Secure,Rouayheb.Soljanin2007}, or the outer code has to be designed based on the specific network code used \cite{Feldman++2004:CapacitySecure}. In all cases, the field size required is significantly larger than the minimum required for conventional multicasting.

The present paper is motivated by Rouayheb and Soljanin's formulation of a wiretap network and builds on their results. Our main contribution is a coset coding scheme that neither imposes any constraints on, nor requires any knowledge of, the underlying network code. In other words, for any linear network code that is feasible for multicast, secure communication at the maximum possible rate can be achieved with a fixed outer code. In particular, the field size can be chosen as the minimum required for multicasting. An important consequence of our result is that the problems of information transport---designing a feasible network code---and security against a wiretapper can be completely separated from each other. Such a feature of our scheme allows it to be seamlessly integrated with random network coding.

The essence of our approach is to use a ``nonlinear'' outer code that is, however, linear over an extension field $\Fqm$. Taking advantage of this extension field, we can then replace the linear MDS code in Ozarow-Wyner coset coding scheme by a maximum-rank-distance (MRD) code, which is essentially a linear code over $\Fqm$ that is optimal in the rank metric. Codes in the rank metric were studied by a number of authors \cite{Gabidulin1985,Roth1991:MaximumRankArrayCodes,Richter.Plass2004:BerlekampMassey,Gadouleau.Yan2006} and have been recently proposed for error control in random network coding \cite{Silva.Kschischang2007:ISIT,Silva++2007:IT}. Here, we show that the fact that the wiretapper observes a linear transformation of the transmitted symbols is exactly what suggests the use of a rank-metric code.

The remainder of the paper is organized as follows. In Section~\ref{sec:wiretap-model} we review the models of a wiretap channel II and a wiretap network, together with their corresponding security conditions. In Section~\ref{sec:rank-metric-for-wiretap} we review rank-metric codes and present our solution to the security problem in a wiretap network. In Section~\ref{sec:discussion}, we provide a brief discussion of our main result and, in Section~\ref{sec:conclusion}, we present our conclusions.

\section{Wiretap Model}
\label{sec:wiretap-model}

\subsection{Wiretap Channel II}
\label{sec:wiretap-channel-II}

Consider a communication system consisting of a source, a destination and a wiretapper. The source produces a message $S = \mat{S_1 & S_2 & \cdots & S_k}^T$, where the symbols $S_1,\ldots,S_k$ are drawn from an alphabet $F$, and encodes this message as a vector $X = \mat{X_1 & \cdots & X_n}^T$, $X_i \in F$. This vector is transmitted over a noiseless channel and received by the destination. The wiretapper has access to $\mu$ symbols of $X$, represented as the vector $W = (X_i,\, i \in \mathcal{I})$, where $\mathcal{I} \subseteq \{1,\ldots,n\}$. The goal of the system is for the source to communicate the message to the destination in such a way that the wiretapper cannot obtain any information about $S$ from any possible set of $\mu$ intercepted symbols. More precisely, the conditions for secure communication are
\begin{align}
H(S|X) &= 0 \label{eq:cond-1} \\
H(S|W) &= H(S),\quad \forall \mathcal{I}\colon |\mathcal{I}|=\mu. \label{eq:cond-2}
\end{align}
Condition (\ref{eq:cond-1}) implies that $S$ must be a deterministic function of $X$. The question is then how to design a (probabilistic) encoding of $S$ into $X$ such that conditions (\ref{eq:cond-1}) and (\ref{eq:cond-2}) are satisfied.

Note that, by expanding $H(S,X|W)$, we have
\begin{align}
H(S|W)
&= \underbrace{H(S|X,W)}_{=0} + H(X|W) - H(X|S,W) \nonumber \\
&= H(X|W) - H(X|S,W) \label{eq:entropy-S-given-W} \\
&\leq H(X|W) \leq n - \mu \nonumber
\end{align}
so the maximum number of symbols that can be securely communicated is upper bounded by $H(S) \leq n-\mu$.

This maximum rate can be achieved by using Ozarow-Wyner coset coding scheme \cite{Ozarow.Wyner1985}, which operates as follows. Assume $F$ is a finite field of sufficiently large cardinality. Let $k = n - \mu$ and let $\mathcal{C}$ be an $(n,\mu)$ linear MDS code over $F$ with parity-check matrix $H$. Encoding is performed by randomly choosing some $X \in \mathcal{C}$ such that $S = HX$; in other words, each message is viewed as a syndrome specifying a coset of $\mathcal{C}$, and the transmitted vector is chosen uniformly at random among the elements of that coset. Upon reception of $X$, decoding is performed by simply computing the syndrome $S = HX$.

With respect to security, it is immediate that condition (\ref{eq:cond-1}) is satisfied in this scheme. Since $\mathcal{C}$ is a linear code, the probabilistic encoding ensures that $H(X) = H(S) + \mu$, and thus $H(X|W) = H(X) - H(W) = H(S) + \mu - H(W) \geq H(S)$. On the other hand, since $\mathcal{C}$ is an MDS code, knowledge of $S$ and $W$ is sufficient to determine $X$, so $H(X|S,W) = 0$. These two facts applied in (\ref{eq:entropy-S-given-W}) imply that condition (\ref{eq:cond-2}) is satisfied, and therefore secure communication can be achieved.

\subsection{Wiretap Networks}
\label{sec:wiretap-networks}

Consider a communication network represented by a directed multigraph with unit capacity edges, a single source node and multiple destination nodes. The source node produces a message $X = \mat{X_1 & \cdots & X_n}^T$ consisting of symbols from an alphabet $F$, and this message is requested by each of the destination nodes. Each link in the network is assumed to transport a symbol in $F$ free of errors. When network coding is used, each node in the network produces symbols to be transmitted by performing arbitrary operations on the received symbols (or on the message symbols in the case of the source node).
We say that the network code is feasible (and multicast communication is achieved) if each destination node is able to recover the source message.

Let $\Fq$ be a finite field and assume that $F$ is a vector space over $\Fq$. In this case, an element of $F$ may also be called a packet. When linear network coding is used, each packet transmitted by a node is an $\Fq$-linear combination of received (or message) packets. Let $C$ be the minimum value of the mincut from the source node to any destination node.
It is a well-known result that a feasible linear network code exists if $n \leq C$ and $q$ is sufficiently large, but no feasible network code exists if $n > C$ \cite{Ahlswede++2000,Li++2003,Koetter.Medard2003}.

The wiretap problem of Section~\ref{sec:wiretap-channel-II} can be generalized to the network scenario above by introducing a wiretapper who can eavesdrop on $\mu$ links, represented by the set $\mathcal{I}$, and by assuming that the source message is given by $S = \mat{S_1 & S_2 & \cdots & S_k}^T$, $S_i \in F$, which is then encoded into $X$ for transmission over the network.
We assume that linear network coding is used, so the packets observed by the wiretapper can be represented as a vector $W = BX$, where $B$ is an $\mu \times n$ matrix over $\Fq$ consisting of the global coding vectors associated with the edges in $\mathcal{I}$.

Assume that $n \leq C$, $q$ is sufficiently large, and that a feasible network code is selected, i.e., each destination node is able to recover $X$. The conditions for secure communication remain the same as before, namely
\begin{align}
H(S|X) &= 0 \label{eq:net-cond-1} \\
H(S|W) &= H(S),\quad \forall \mathcal{I}\colon |\mathcal{I}| = \mu. \label{eq:net-cond-2}
\end{align}
The question is then how to design an encoding from $S$ to $X$ \emph{and} a feasible linear network code such that (\ref{eq:net-cond-1}) and (\ref{eq:net-cond-2}) are satisfied.

Considering $F = \Fq$, Rouayheb and Soljanin showed in \cite{Rouayheb.Soljanin2007} that secure communication is possible using the coset coding scheme of Sec.~\ref{sec:wiretap-channel-II} if the network code is chosen to satisfy certain constraints. The development is similar to that of Sec.~\ref{sec:wiretap-channel-II}, where we choose $k = n - \mu$ and let $H$ be the parity-check matrix of an $(n,\mu)$ linear MDS code over $F$. Equations (\ref{eq:net-cond-1}) and $H(X|W) \geq H(S)$ are automatically satisfied by coset encoding, but to satisfy $H(X|S,W) = 0$ we must ensure that
%the system of equations
%\begin{align}
%S &= HX \nonumber \\
%W &= BX \nonumber
%\end{align}
%has a unique solution in $X$.
the matrix
%\begin{equation}\nonumber
  $\mat{H \\ B}$
%\end{equation}
is nonsingular for all $\mathcal{I}$ such that $B$ is full-rank. (Note that the case where $B$ is not full-rank reduces to a similar instance with a full-rank $B$ and a smaller $\mu$.) This condition is equivalent to constraining the network code such that no linear combination of $\mu = n-k$ or fewer coding vectors belongs to the space spanned by the rows of $H$.

It follows from this result that secure multicast communication can be achieved in two steps: first, designing a coset coding scheme based on an MDS code, and then designing a linear network code so as to satisfy the above constraint.

In the following, we show that this undesirable coupling between the coset coding scheme and the network code design can be avoided.

\section{Rank-Metric Codes for Wiretap Networks}
\label{sec:rank-metric-for-wiretap}

\subsection{Rank-Metric Codes}
\label{sec:rank-metric-codes}

We first present a brief review of rank-metric codes.

Let $\Fq^{n \times m}$ be the set of all $n \times m$ matrices over $\Fq$. A natural distance measure between elements $X$ and $Y$ of $\Fq^{n \times m}$ is given by the \emph{rank distance} $d_R(X,Y) \triangleq \rank(Y - X)$. As observed in \cite{Gabidulin1985}, the rank distance is indeed a metric.

A \defn{rank-metric code} is a nonempty subset of $\Fq^{n \times m}$ used in the context of the rank metric. The minimum rank distance of a rank-metric code is the minimum rank distance among all pairs of distinct codewords.
%\begin{equation}\nonumber
%  D_R(\mathcal{C}) \triangleq \min_{\substack{X,X' \in \mathcal{C} \\ X \neq X'}} d_R(X,X').
%\end{equation}
The Singleton bound for the rank metric (see \cite{Silva++2007:IT,Gadouleau.Yan2006} and references therein) states that every rank-metric code $\mathcal{C} \subseteq \Fq^{n \times m}$ with minimum rank distance $d$ must satisfy
\begin{equation}\nonumber
\log_q |\mathcal{C}| \leq \max\{n,m\} (\min\{n,m\} - d + 1).
\end{equation}
Codes that achieve this bound are called \defn{maximum-rank-distance} (MRD) codes.

The usual way to construct rank-metric codes is via the correspondence between $\Fq^{1 \times m}$ and an extension field $\Fqm$. By fixing a basis for $\Fqm$ as an $m$-dimensional vector space over $\Fq$, any element of $\Fqm$ can be regarded as a \emph{row} vector of length $m$ over $\Fq$ and, similarly, any \emph{column} vector of length $n$ over $\Fqm$ can be regarded as an $n \times m$ matrix over $\Fq$. The rank of a vector $X \in \Fqm^n$ is the rank of $X$ as an $n \times m$ matrix over $\Fq$, and the same applies for the rank distance. Under this correspondence, a rank-metric code in $\Fq^{n \times m}$ is simply a block code of length $n$ over $\Fqm$ used in the context of the rank metric.

It is useful to consider \emph{linear} $(n,k)$ codes over $\Fqm$ with minimum rank distance $d$. For such codes, the Singleton bound becomes
\begin{equation}\nonumber
  d \leq \min\left\{1,\frac{m}{n}\right\} (n-k) + 1.
\end{equation}
Note that the classical Singleton bound $d \leq n - k + 1$ can be achieved only when $n \leq m$. For this case, a class of MRD codes with any specified $k$ was described in \cite{Gabidulin1985} by Gabidulin.

We now restate some results from \cite{Gabidulin1985} which relate the minimum rank distance of a linear code with properties of its parity-check matrix. To avoid confusion, the rank of a matrix $H$ over $\Fqm$ is denoted by $\rank_{q^m} H$.

\medskip
\begin{theorem}
  Let $\mathcal{C}$ be a linear $(n,k)$ code over $\Fqm$ with parity-check matrix $H$. Then $\mathcal{C}$ has minimum rank distance $d$ if and only if
  \begin{equation}\nonumber
  \rank_{q^m} HT = d-1
  \end{equation}
  for any full-rank matrix $T \in \Fq^{n \times (d-1)}$ and
  \begin{equation}\nonumber
  \rank_{q^m} HT_0 < d
  \end{equation}
  for some full-rank matrix $T_0 \in \Fq^{n \times d}$.
\end{theorem}
\medskip

\begin{corollary}\label{cor:rank-H-MRD}
  Assume $n \leq m$. A linear $(n,k)$ code over $\Fqm$ with parity-check matrix $H$ is an MRD code if and only if
  \begin{equation}\nonumber
  \rank_{q^m} HT = n - k
  \end{equation}
  for any full-rank matrix $T \in \Fq^{n \times (n-k)}$.
\end{corollary}

\subsection{A Universal Coding Scheme for Wiretap Networks}
%\label{sec:}

We now present our solution to the wiretap problem of Section~\ref{sec:wiretap-networks}. Following \cite{Rouayheb.Soljanin2007}, we use a coset coding scheme similar to that of Section~\ref{sec:wiretap-channel-II}; however, we set the symbol alphabet to be $F = \Fqm$, while the field for the linear network coding operations remains $\Fq$. Note that, since coset encoding/decoding is performed only at source/destination nodes, setting $F$ to be an extension field of $\Fq$ does not interfere with the underlying network code.

Let $k = n - \mu$ and let $H$ be the parity-check matrix of a linear $(n,\mu)$ code over $F$. Encoding and decoding of the source message $S$ is performed as described in Section~\ref{sec:wiretap-channel-II}. With respect to security, Rouayheb and Soljanin's analysis carries out unchanged, and we arrive at the same security condition: the matrix
%\begin{equation}\nonumber
  $\mat{H \\ B}$
%\end{equation}
must be nonsingular for all $\mathcal{I}$ such that $B \in \Fq^{\mu \times n}$ is full-rank.
Note that, while $H$ is defined over $F = \Fqm$, the matrix $B$ has only entries in $\Fq$. This fact is the fundamental distinction of our approach and will allow us to satisfy the security condition regardless of the network code used.

Our main result is a consequence of the following lemma.

\medskip
\begin{lemma}
  Let $H$ be the parity-check matrix of a linear MRD $(n,\mu)$ code over $\Fqm$. For any full-rank matrix $B \in \Fq^{\mu \times n}$, the $n \times n$ matrix
  \begin{equation}\nonumber
    M = \mat{H \\ B}
  \end{equation}
  is nonsingular over $\Fqm$.
\end{lemma}
\begin{IEEEproof}
Consider the system of equations
\begin{equation}\nonumber
  \mat{H \\ B} X = 0
\end{equation}
in the unknown $X \in \Fqm^n$. We will show that $X=0$ is the only solution to this system, which implies that $\rank_{q^m} M = n$.

First, choose some $(n-\mu) \times n$ matrix $D$ over $\Fq$ such that $\mat{B \\ D}$ is nonsingular, and let $\tilde{X} = DX$. We have that
\begin{equation}\nonumber
  \mat{B \\ D}X = \mat{0 \\ \tilde{X}} \implies X = \mat{B \\ D}^{-1}\mat{0 \\ \tilde{X}}.
\end{equation}
Moreover, if $T$ is the (full-rank) matrix corresponding to the last $n-\mu$ columns of $\mat{B \\ D}^{-1}$, then $X = T \tilde{X}$.

Now, $0 = HX = HT \tilde{X}$. By Corollary~\ref{cor:rank-H-MRD}, the $(n-\mu) \times (n-\mu)$ matrix $HT$ is nonsingular over $\Fqm$. Thus, we must have $\tilde{X} = 0$ and hence $X = 0$.
\end{IEEEproof}
\medskip

The following theorem summarizes the results of this section.

\medskip
\begin{theorem}\label{thm:main-result}
Consider a multicast communication network that transports $n$ packets of length $m \geq n$ over $\Fq$, subject to the presence of a wiretapper who can eavesdrop on at most $\mu$ links.
The maximum number of source packets that can be securely communicated to each destination, in such a way that the wiretapper obtains no information about the source packets, is $n - \mu$. This rate can be achieved by using \emph{any} feasible $\Fq$-linear network code in conjunction with a \emph{fixed} end-to-end coset coding scheme based on any linear MRD $(n,\mu)$ code over $\Fqm$.
%For \emph{any} feasible $\Fq$-linear network code, secure communication of up to $n-\mu$ packets can be achieved using a \emph{fixed} coset coding scheme based on a linear MRD $(n,\mu)$ code over $\Fqm$. Conversely, if $k > n - \mu$, then the uncertainty that the wiretapper has about the source packets will be smaller than $k$.
\end{theorem}
\medskip

The following example illustrates the above results.

\medskip
\begin{example}
  Let $q=2$, $m=n=3$, $\mu=2$ and $k=n-\mu=1$. Let $F = \mathbb{F}_{2^3}$ be generated by a root of $p(x) = x^3 + x + 1$, which we denote by $\alpha$. According to \cite{Gabidulin1985}, one possible $(n,\mu)$ MRD code over $\Fqm$ has parity-check matrix $H = \mat{1 & \alpha & \alpha^2}$.

To form $X$, we can choose $X_2,X_3 \in \Fqm$ uniformly at random and set $X_1$ to satisfy
\begin{equation}\nonumber
  S = HX = X_1 + \alpha X_2 + \alpha^2 X_3.
\end{equation}
Note that $X$ can be transmitted over any network that uses a feasible linear network code. The specific network code used is irrelevant as long as each destination node is able to recover $X$.

Now, suppose that the wiretapper intercepts $W = BX$, where
\begin{equation}\nonumber
  B = \mat{1 & 0 & 1 \\ 0 & 1 & 1}.
\end{equation}
Then
\begin{align}
  W &= B\mat{X_1 \\ X_2 \\ X_3} = \mat{1 & 0 & 1 \\ 0 & 1 & 1}\mat{S + \alpha X_2 + \alpha^2 X_3 \\ X_2 \\ X_3} \nonumber \\
&= \mat{1 \\ 0}S +  \mat{\alpha & 1+\alpha^2 \\ 1 & 1} \mat{X_2 \\ X_3}. \nonumber
\end{align}
This is a linear system with $3$ variables and $2$ equations over $\Fqm$. Note that, given $S$, there is exactly one solution for $(X_2,X_3)$ for each value of $W$. Thus, ${\sf Pr}(W|S) = 1/8^{2}$, $\forall S,W$, from which follows that $S$ and $W$ are independent.
\end{example}

\section{Discussion}
\label{sec:discussion}

Theorem~\ref{thm:main-result} shows that the problem of ensuring communication security against a wiretapper can be treated independently from that of multicasting information, in effect turning network coding design back into a much easier and already satisfactorily solved problem \cite{Jaggi++2005:PolynomialTimeAlgorithms}. A byproduct of this result is that, to incorporate security, we no longer need to enlarge the field of network coding operations more than what is strictly required for multicasting---although the network does need to transport packets of size larger than a single element.
% (in fact, larger than the total number of packets transmitted).
In practice, packet lengths are much larger than $n$, at least 10 times larger for typical parameters, so the constraint $m \geq n$ is not really a concern.

As pointed out in the previous section, encoding and decoding of the source message require operations to be performed in the extension field $\Fqm$. We mention that each encoding or decoding procedure can be performed in $\bigO(k(n-k))$ operations in $\Fqm$ by using a parity-check matrix $H$ in systematic form. More precisely, if $H = \mat{I & P}$ and $X^T = \mat{X_S^T & X_R^T}$, where $X_S$ has $k$ rows, then
%\begin{equation}\nonumber
  $S = HX = X_S + PX_R$,
%\end{equation}
so $S$ can be encoded by randomly generating $X_R$ and then setting $X_S = S - PX_R$. Encoding thus amounts essentially to a matrix multiplication over $\Fqm$. Decoding can be performed similarly.
%Since each multiplication in $\Fqm$ can be performed in $\bigO(m^2)$ operations in $\Fq$, the overhead per transmitted symbol (assuming $m=n$) is proportional to $k(n-k)$.

It is worth to mention that our security scheme can be seamlessly integrated with random network coding. We simply require that each packet transports a header of length $n$ containing the global coding vector associated with the packet; thus, the total packet length must be at least $n+m$ symbols in $\Fq$. Note that, since a random linear network code is feasible with high probability, the only parameter pertaining to the network that we need to estimate is the effective mincut $C$, in order to decide on $n$, $k$ and the coset coding scheme.

\section{Conclusion}
\label{sec:conclusion}

We consider the problem of providing information-theoretic security in a communication network subject to the presence of a wiretapper. We propose a coset coding scheme similar to that of Ozarow-Wyner, but defined over the extension field $\Fqm$. For this reason, we assume that packets of length $m$ are transmitted rather than individual symbols. We show that transmission at the maximum possible rate (the network secure capacity) is possible irrespectively of the underlying network code. As a consequence, the sub-problems of information transport and information security can be treated independently of each other: a feasible linear network code can be designed (perhaps, randomly) with only throughput in mind, while a \emph{fixed} outer code can be used to provide security whenever it is needed. Our proposed scheme is based on MRD codes and can be efficiently encoded and decoded.

\bibliographystyle{IEEEtran}
\bibliography{networkcoding,codingtheory,rankmetric}

\end{document}